\documentclass[sigconf]{acmart}

\usepackage{booktabs}
\usepackage{multirow}
\usepackage{times}
\usepackage{epsfig}
\usepackage{graphicx}
\usepackage{amsmath}
\usepackage{amssymb}
\usepackage{caption}
\usepackage{subcaption}
\usepackage{tabularx}

\hypersetup{
	colorlinks = true,
	citecolor = blue
}

\makeatletter
\renewcommand\@makefnmark{\hbox{\@textsuperscript{\normalfont\color{red}\@thefnmark}}}
\makeatother

\begin{document}
\title{An Ensemble-Based Approach to Click-Through Rate Prediction for Promoted Listings at Etsy}
\author{Kamelia Aryafar}
\affiliation{%
  \institution{Etsy}
  \streetaddress{117 Adams Street}
  \city{Brooklyn} 
  \state{NY} 
}
\email{karyafar@etsy.com}
\author{Devin Guillory}
\affiliation{%
    \institution{Etsy}
  \streetaddress{20 California Street}
  \city{San Francisco} 
  \state{CA} 
}
\email{dguillory@etsy.com}
\author{Liangjie Hong}
\affiliation{%
  \institution{Etsy}
  \streetaddress{117 Adams Street}
  \city{Brooklyn} 
  \state{NY} 
}
\email{lhong@etsy.com}

\copyrightyear{2017} 
\acmYear{2017} 
\setcopyright{acmcopyright}
\acmConference{ADKDD'17}{August 14, 2017}{Halifax, NS, Canada}\acmPrice{15.00}\acmDOI{10.1145/3124749.3124758}
\acmISBN{978-1-4503-5194-2/17/08}

\begin{abstract}
Etsy~\footnote{\url{http://www.etsy.com}} is a global marketplace where people across the world connect to make, buy, and sell unique goods. Sellers at Etsy can promote their product listings via advertising campaigns similar to traditional sponsored search ads. Click-Through Rate ({\tt CTR}) prediction is an integral part of online search advertising systems where it is utilized as an input to auctions which determine the final ranking of promoted listings to a particular user for each query. In this paper, we provide a holistic view of Etsy's promoted listings' {\tt CTR} prediction system and propose an ensemble learning approach which is based on historical or behavioral signals for older listings, as well as content-based features for new listings. We obtain representations from texts and images by utilizing state-of-the-art deep learning techniques and employ multimodal learning to combine these different signals. We compare the system to non-trivial baselines on a large-scale, real world dataset from Etsy, demonstrating the effectiveness of the model and strong correlations between offline experiments and online performance. The paper is also the first technical overview to this kind of product in an e-commerce context.
\end{abstract}

%
%

\ccsdesc[500]{Information systems~Learning to rank}
\ccsdesc[300]{Information systems~Evaluation of retrieval results}
\ccsdesc[300]{Computing methodologies~Neural networks}
\ccsdesc[300]{Computing methodologies~Ensemble methods}

\keywords{Click-Through Rate Prediction, Logistic Regression, Multimodal Learning, Deep Learning}

\maketitle

\section{Introduction}\label{sec:introduction}

\begin{figure}
	\centering
	\includegraphics[scale=0.2]{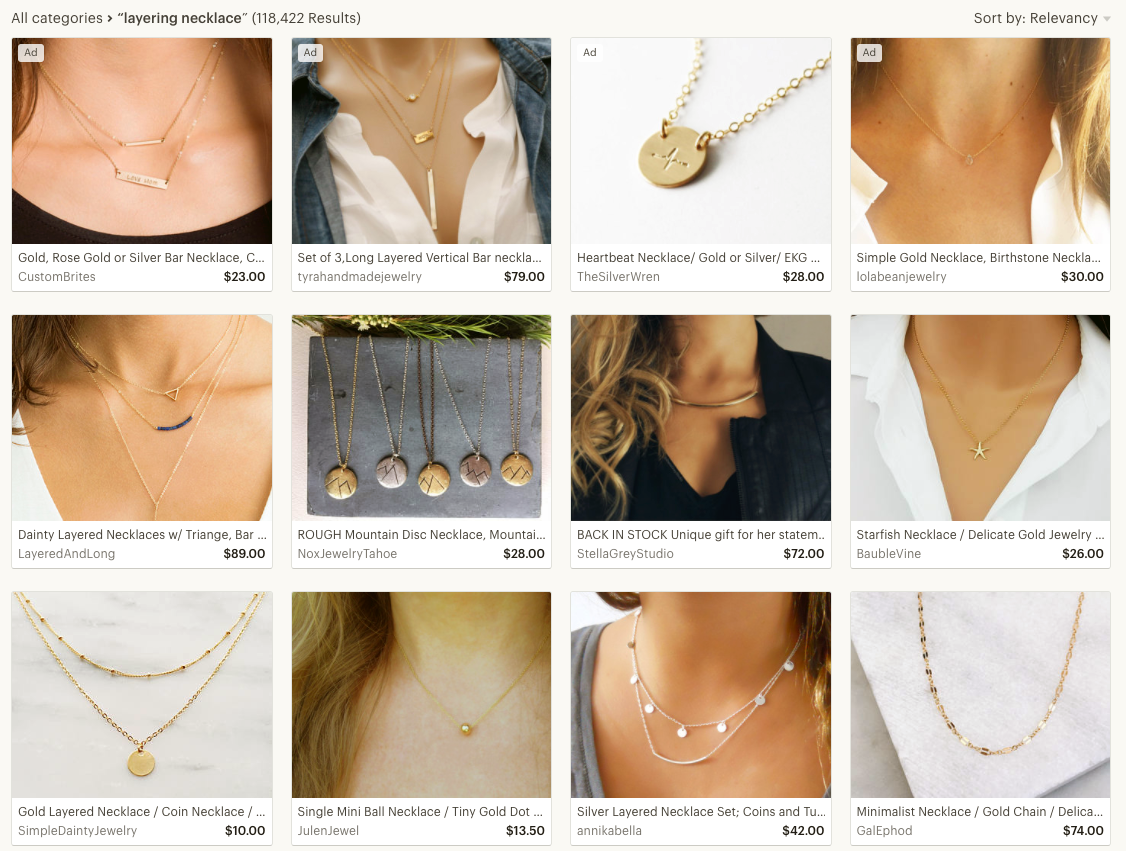}
	\caption{Promoted and organic search results for a query {\tt layering necklace} is shown where promoted listings are presented in the first row and organic results followed.}
	\label{fig:motivation}
\end{figure}

Etsy, founded in 2005, is a global marketplace where people around the world connect to make, buy, and sell unique goods, including handmade items, vintage goods, and craft supplies. Users come to Etsy to search for and buy listings other users offer for sale. Currently, Etsy has more than 45M items to sell with nearly 2M active sellers and 30M active buyers across the globe\footnote{\url{https://www.etsy.com/about/}}. As millions of people search for items on the site every day and increasingly more sellers compete for more customers, Etsy has started to offer a \textit{Promoted Listings Program}\footnote{\url{https://www.etsy.com/advertising\#promoted\_listings}} to help sellers get their items in front of these interested shoppers by placing some of their listings higher up in related search results. An example of promoted listings, blending with organic results, is shown in Figure \ref{fig:motivation}.

In a nutshell, promoted listings at Etsy work as follows. A seller who is willing to participate in the program would specify a total budget that he/she wants to spend during the whole campaign. Etsy, as the platform, would choose queries or keywords that the campaign runs on and how much to bid for each query. In a simplified setting, for each query $q$ and each relevant promoted listing $l$, the platform computes $b_{l,q}$, the bidding price of the listing $l$ to the query $q$, and the expected Click-Through Rate ({\tt CTR}) $\theta_{l,q}$. The score $b_{l,q} \times \theta_{l,q}$ is used in a \textit{generalized second price auction}\footnote{\url{https://en.wikipedia.org/wiki/Generalized\_second-price\_auction}} with other qualified listings. The final pricing and the position of a promoted listing is determined by the auction. The whole process is similar to traditional sponsored search with two distinctions: 
\begin{itemize}
	\item Sellers can only specify the overall budget without the control on each bidding price.
	\item Sellers cannot choose which queries they want to bid on.
\end{itemize}For both aspects, Etsy would operate on sellers' behalf. Etsy uses a Cost-Per-Click ({\tt CPC}) model, meaning that the site charges sellers when a buyer clicks on the promoted listing. A similar program exists in eBay\footnote{\url{http://pages.ebay.com/seller-center/stores/promoted-listings/benefits.html}} although it uses Cost-Per-Action ({\tt CPA}) model. As Etsy is using {\tt CPC} model to operate promoted listings, one way to optimize the platform's revenue is to increase the number of clicks for each promoted listing. In other words, we would like to have higher $b_{l,q}$ and $\theta_{l,q}$ for each clicked listing $l$ to the query $q$. 

In this paper, we discuss the methodologies and systems to drive {\tt CTR} $\theta$, given a fixed bidding strategy which computes $b$. To our knowledge, our paper is the first study to systematically discuss how a promoted listings system can be built with practical considerations. In particular, we focus on areas to address the following research and engineering questions:
\begin{itemize}
	\item What is the right architecture to balance the effectiveness and simplicity of the system, given the current scale of Etsy?
	\item What are effective {\tt CTR} prediction algorithms, features, and modeling techniques?
	\item As an e-commerce site where images are ubiquitous and vital to the user experience, what is the appropriate strategy to incorporate such information into the algorithm?
	\item Is there any correlation between offline evaluation metrics and online model performance, such that we can constantly improve our models?
\end{itemize}Although some of these questions might have been tackled in prior work (\textit{e.g.}, \cite{McMahan2013,He2014,Agarwal2014,Li2015}), we provide a comprehensive view of these issues in this paper. In addition to the holistic view, in terms of modeling techniques, we propose an ensemble learning approach which leverages users' behavioral signals for existing listings and content-based features for new listings. In order to learn meaningful representations from features, we utilize feature hashing and deep learning techniques to obtain representations from texts and images and employ a multimodal learning process to combine these different signals. To demonstrate the effectiveness of the system, we compare the proposed system to non-trivial baselines on a large-scale, real world dataset from Etsy with offline and online experimental results. During this process, we also establish a strong correlation between offline evaluation metrics and online performance ones, which serves as a guidance for future experimental plans.

The paper is organized as follows. We start with a brief review of related work in Section \ref{sec:related}, followed by the discussion of our methodology in Section \ref{sec:method}. We demonstrate the effectiveness of the system in Section \ref{sec:experiments} and conclude the paper in Section \ref{sec:conclusion}.


\section{Related Work}\label{sec:related}
In this section, we briefly review some related work to online advertising and multimodal modeling with respect to learning to rank.

{\bfseries Ads' CTR Prediction}: Several industrial companies shared their practical lessons and methodologies to build large-scale ads systems. Early authors from Microsoft proposed online Bayesian probit regression ~\cite{Graepel2010} to tackle the {\tt CTR} prediction problem. Arguably the first comprehensive review is from Google's ads system~\cite{McMahan2013} where the paper not only talks about different aspects of machine learning algorithms (\textit{e.g.}, models, features, and calibration techniques.) to improve ads click-through-rate modeling, but it also lays out what kind of components need to be built for a robust industrial system (\textit{e.g.}, monitoring and alerts). The paper also populated Follow-The-Regularized-Leader ({\tt FTRL}) as a strong ads {\tt CTR} prediction baseline. In~\cite{He2014}, authors from Facebook described their ads system with the novelty of utilizing Gradient Boosted Decision Trees ({\tt GBDT}) as feature transformers, as well as online model updates and data joining. Twitter's engineers and researchers~\cite{Li2015} advanced the field by applying both pointwise and pairwise learning to {\tt CTR} prediction problem with detailed feature analysis. LinkedIn's authors~\cite{Agarwal2014} proposed to utilize ADMM to fit large-scale Bayesian logistic regression and also provided detailed discussion on how to cache model coefficients and features. Chapelle et al.~\cite{Chapelle2014} described Yahoo's {\tt CTR} prediction algorithms with the emphasis on a wide range of techniques to improve the accuracy of the model including subsampling, feature hashing, and multi-task formalism.

{\bfseries Multimodal Learning to Rank}: There exists many works to utilize multimodal data, especially images, for the scenario of learning to rank. The basic idea behind multimodal learning is to learn different representations from different data types and combine their predictive strengths in another layer (\textit{e.g.}, \cite{Andrew2013,Kiros2014}). Specifically in learning to rank, Lynch et al.~\cite{Lynch2016} utilized deep neural networks to extract images features, and learned a pairwise {\tt SVM} model to rank organic search results to queries. Recently proposed {\tt ResNet}~\cite{He2016} has been widely treated as a generic image feature generation tool, which is used in the current work.


\section{Systems and Methodologies}\label{sec:method}
On a high level,  our promoted listings system matches a user query with an initial set of promoted listing results based on the textual relevance. The final ranking of the listings is determined by a generalized second price auction mechanism that takes multiple factors such as bids, budgets, pacing, and {\tt CTR} of those items into account. In this paper, we focus on the {\tt CTR} prediction of the system.

\subsection{System Overview}\label{sec:method_system_overview}
\begin{figure}
	\centering
	\includegraphics[scale=0.35]{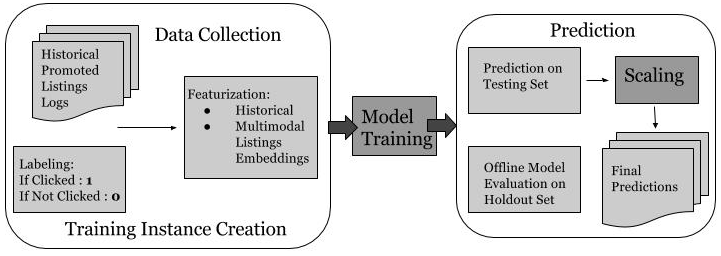}
	\caption{The high-level {\tt CTR} prediction system overview, which is described in Section~\ref{sec:method_system_overview} and historical and multimodal listings embedding is covered in Section~\ref{sec:method_features}, model training in Section~\ref{sec:method_ctr_prediction}, offline evaluations and predictions in Section~\ref{sec:exp_metrics} and Section~\ref{sec:exp_res}}
	\label{fig:sysarch}
\end{figure}
Our {\tt CTR} prediction system consists of three main components:
\begin{enumerate}
	\item data collection and training instances creation;
	\item model training and deployment;
	\item inference and {\tt CTR} prediction scores serving.
\end{enumerate}First, real-time events, including which promoted listings are served and how users interact with them, are processed through our Kakfa~\footnote{\url{https://kafka.apache.org/}} pipeline to our Hadoop Distributed File System (HDFS) regularly. We then extract listings' information with historical user behavior data from HDFS. During this process, we assign positive and negative labels to listings where each clicked listing is considered as a positive data instance while each non-clicked listing is considered as negative. The listings information is transformed into a feature vector containing both historical and multimodal embeddings described in Section~\ref{sec:method_features}. All listings features, excluding the image representations, are extracted via multiple MapReduce Hadoop jobs. The deep visual semantic features are extracted in a batch mode on a set of single boxes with GPUs using Torch~\footnote{\url{http://torch.ch/}}. The visual features are then transferred to HDFS where they are joined with all other listing features to create a labeled training instance.

We utilize Vowpal Wabbit~\footnote{\url{http://hunch.net/~vw/}} to train our {\tt CTR} prediction models where all training processes happen on a single box. The details about {\tt CTR} prediction model are described in Section~\ref{sec:method_ctr_prediction}. The model is then deployed to multiple boxes as an API endpoint for inference. The new model is validated through a set of offline evaluation metrics discussed in Section~\ref{sec:exp_metrics} on a holdout dataset of promoted listings search logs.

At inference time, a new {\tt CTR} prediction score is generated for every active listing available for sale. These listings are tagged with the feature representations similar to training side via multiple MapReduce Hadoop jobs to create testing instances. The model is applied on each testing instance via the API endpoint in a MapReduce job where a {\tt CTR} prediction score is returned. The {\tt CTR} prediction scores are calibrated to control for {\tt CPC} by keeping the {\tt CTR} mean and standard deviation the same across multiple experimental variants. The calibrated {\tt CTR} prediction is the final input to the auction mechanism deciding the final ranking of promoted listings. Figure~\ref{fig:sysarch} provides an overview of this architecture.

\subsection{CTR Prediction}\label{sec:method_ctr_prediction}
In a nutshell, given a query $q$ issued by a user, {\tt CTR} prediction determines how likely a listing $l$ is going to be clicked by the user under the context of $q$. In other words, the system models the conditional probability $P(c \, | \, q, l)$ where $c$ is the binary response, click or not. 

In practice, we form a feature vector $\boldsymbol{x}_{q,l} \in \mathbb{R}^{d}$ to represent the overall feature representation of query $q$ and listing $l$. In the latter part of the paper, we will discuss how $\boldsymbol{x}_{q,l}$ is constructed in detail. Since logistic regression is a widely-used \textit{de-facto} machine learning model for {\tt CTR} prediction in industry due to its scalability and speed, we also follow the same formalism, modeling $P(c \, | \, q, l) = \sigma(\boldsymbol{w} \cdot \boldsymbol{x}_{q,l})$ where $\boldsymbol{w} \in \mathbb{R}^{d}$ is a coefficient vector and $\sigma(a) = 1/(1+\exp(-a))$ is a sigmoid function. Many approaches exist to obtain optimal $\boldsymbol{w}$. Here, we utilize {\tt FTRL-Proximal} algorithm as the primary learning tool for both single and multimodal feature representations discussed in Section~\ref{sec:method_features}:
\begin{align*}
\mathbf{w}_{t+1} = \arg\min_{\mathbf{w}} \bigl( \mathbf{g}_{1:t} \cdot \mathbf{w} + \frac{1}{2} \sum_{s=1}^{t}\sigma_{s} || \mathbf{w} - \mathbf{w}_{s}||^{2}_{2} + \lambda_{1} || \mathbf{w} ||_{1}\bigr)
\end{align*}where $\mathbf{w}_{t+1}$ is the optimal coefficient vector at iteration $t+1$, $\mathbf{g}_{1:t}=\sum_{s=1}^{t} \mathbf{g}_{s}$ ($\mathbf{g}_{s}$ are the gradients at iteration $s$) and $\sigma_{s}$ is the learning-rate schedule and $\lambda_{1}$ is the regularization parameter. The detailed description of the algorithm is discussed in~\cite{McMahan2013}.

\subsection{Feature Representations}\label{sec:method_features}
The feature representation $\boldsymbol{x}$ plays a vital role in the effectiveness of the model. Here, we focus on the sets of listing features that are important to the performance. The listing features used in {\tt CTR} prediction can be divided into two sets of features: \textit{historical} features based on promoted listing search logs that record how users interact with each listing and \textit{content-based} features, which are extracted from the information presented in each listing's page.


{\bfseries Historical Features}: This type of feature tracks how a particular listing performs in terms of {\tt CTR} and other behavior metrics in the past, which is a usually strong baseline to the problem. Here, we describe how smoothed historical {\tt CTR} works. Other types of historical features can be easily extended from this discussion. 

For a listing $l_{i}$, we assume that click events are random variables drawn from a Binomial distribution with parameter $\theta_{i}$ being the probability to be clicked by users. The na\"{\i}ve estimator is also  the Maximum Likelihood Estimator ({\tt MLE}), $\hat{\theta}_{i} = \frac{c_{i}}{v_{i}}$ where $c_{i}$ is the number of clicks for the item $i$ and $v_{i}$ is the number of impressions for the item $i$. This estimator is only reliable for listings with significant enough data. However, this is usually not the case for new listings with much less or zero impressions. This motivates us to use a prior distribution to smooth the estimation, similar to the method described in Section 3.1 of ~\cite{Agarwal2009}. Essentially, we put a $\mbox{Beta}(\alpha, \beta)$ prior on $\theta_{i}$ and the mean of the posterior distribution (which is also a Beta distribution due to conjugacy) becomes:
\begin{align*}
\frac{c_{i} + \alpha}{v_{i} + \alpha + \beta}
\end{align*}where $\alpha$ can be set as the global average of clicks and $\beta$ can be set as the global average of impressions (We understand that there exists empirical Bayes method to estimate both parameters from data). We denote this estimator as \textit{smoothed} {\tt CTR}, which has a smaller variance and is more stable compared to the {\tt MLE} estimator. In practice, we re-compute the average number of clicks and impressions for $\alpha$ and $\beta$ for every time period and compute the exponential smoothing over these numbers to make sure the prior numbers are stable. The smoothed {\tt CTR} is computed over the training set per day and is used as the only feature for a logistic regression model, denoted as the \textit{historical} model discussed in Section~\ref{sec:exp_res}. From our experiments, we found that this model serves a strong baseline and smoothed {\tt CTR} also contributes significantly in the final predictive model.

In addition to historical {\tt CTR}, we compute similar historical features for other user behaviors, including listing favorites, purchases, and etc.    


{\bfseries Content-Based Features}: Etsy listings are composed of text information such as tags, title and description, numerical information such as price and listing ID and at least one main image representing the item for sale. The goal here is to learn an effective multimodal feature representation $F_{l}$ for a listing item $l$ given its title, tag words, a numerical listing ID, a numerical price, and a listing image.

For text, we obtain a feature vector by embedding unigrams and bigrams of tags and title in a text vector space, using the hashing trick~\cite{Weinberger2009}. Since raw text can grow unbounded and most raw features only appear handful of times, the feature hashing technique can help us have a finite and meaningful representation of texts. We denote the text embedding as $T_{l}$. 

For images, we embed each image into a representation denoted by $I_{l}$. Here, $I_{l}$ is obtained by adopting a transfer learning method~\cite{Oquab2014} where the internal layers of a pre-trained convolutional neural network act as generic feature extractors on other datasets with or without fine tuning depending on the specs of the dataset. The image features are transfer-learned from a pre-trained deep residual neural network (ResNet101~\cite{He2016}) that is pre-trained on ImageNet~\cite{Deng2009}. We utilize the last fully connected layer of the residual neural network to obtain a $20:48$ dimensional representation of listing $l$ in the image space.

The multimodal embedding of the item $l$ is then obtained as $F_ {l} = [T_{l}, I_{l}]$, which is a simple concatenation of both text embedding features and the image embedding features. The final multimodal feature representation is used to train the logistic regression (dented as \textit{content-based} model) discussed in Section~\ref{sec:exp_res}.

\subsection{Ensemble Learning}\label{sec:method_ensemble}
As we discussed in \ref{sec:method_features}, different features might capture different aspects of listings such that it may not perform well by simply concatenating them together. From our preliminary experiments, we found that historical features and listing embedding features are effective in slightly different ways. Historical features show strong performance for listings with enough data but perform poorly for the ones with little data or no data (behaving like prior distribution in those cases). On the other hand, content-based features demonstrate robust performance for listings with little data. Based on these observations, we propose an ensemble learning approach to combine these different features. To be specific, we train separate {\tt CTR} prediction models purely based on historical features and content-based features. Then, we combine these individual models with a higher level model, again another logistic regression model. This higher level model determines how to balance signals (in this case, prediction scores) from models based on historical features and models based on content-based features. In practice, we train multiple such higher level models with different regions of data (\textit{e.g.}, listings with enough historical data or not), which will be discussed in Section \ref{sec:experiments}.


\section{Experimental Setup}\label{sec:experiments}
In this section, we present the results of our {\tt CTR} prediction system on a number of experiments on real-world, large scale datasets from Etsy's online marketplace. We follow the common practice of \textit{Progressive Validation}~\cite{McMahan2013}, where for each modeling pipeline we have the training data extracted from $[t-32, t-2]$ and $t$ is the current date. The date $t-1$'s data is used as a validation set for model comparisons and parameter tuning. The winning model can be deployed on date $t$. We can run multiple such pipelines to support different variants for online A/B testing. Therefore, under this framework, we can consistently innovate on models and compare offline and online model performances.

Here, we report a set of experimental results from one date. The dataset for the date consists of more than $35$ million Etsy listings. We collect the training data from $[t-32, t-2]$ days of promoted listing search logs across all users. Each training instance is an active listing with an impression within the past $30$ days. We assign positive labels to clicked listings and negative ones to listings with no clicks. Since the dataset is heavily imbalanced with non-clicked impressions dominating, we subsample the negative class where the down-sampled training dataset consists of $26.5$ million data instances. The $t-1$'s validation set consists of $19.5$ million instances that are collected over one day of promoted search logs. We report the offline evaluation metrics on this dataset which is in line with the live A/B tests, as discussed in section~\ref{sec:exp_metrics}.


\subsection{Evaluation Metrics}\label{sec:exp_metrics}
Meaningful offline evaluation metrics that establish correlations with online A/B test results are a key requirement of production quality iterative systems such as {\tt CTR} prediction systems. Online A/B tests can be expensive in a number of ways, especially for a particular web service in which the overall traffic is fixed for a given period of time and therefore usually cannot afford to run a large number of experiments. Each experiment requires at least several days to weeks, sometimes months, to tell the statistically significant difference between the control and the treatment group. Thus, it becomes critical to launch experiments with confidence. In other words, we want to launch models which win experiments with high probability and reduce the cost of running many online A/B tests. In order to achieve this, we would like to seek correlations between offline experimental metrics and online A/B testing metrics such that we could use offline metrics to determine which variant would likely make a successful online A/B candidate. Similar ideas have been also explored in ~\cite{Yi2013}.

The overall process to establish the correlation between offline metrics and online metrics is complicated. Here, we describe a simplified version. In short, we launched a number of experiments with \textbf{known} offline effects (e.g., a particular model outperforms/underperforms the baseline production model in a number of offline metrics which are described below). Then, we observed how these models perform in online A/B tests and saw among all these offline metrics, which ones are good indicators for online key business metrics (e.g., clicks, revenue, etc.). In general, we not only wanted to seek a sign-correlation (e.g., an offline win indicates an online win or an offline lose implies an online lose) but also wanted to have the right magnitude (e.g., single digit  {\tt AUC} win indicates single digit {\tt CTR} win for A/B tests). It should be noted that through multiple such studies, we have established \textit{the \textbf{A}rea \textbf{U}nder the Receiver Operating Characteristic \textbf{C}urve ({\tt AUC})} as a reliable indicator of predictive power of our models.

In our system, we monitor  {\tt AUC}, \textit{Average Impression Log Loss} and \textit{Normalized Cross Entropy}~\cite{Yi2013} as key metrics, which are widely used in industry as offline evaluation metrics. Here, we discuss these three metrics in detail.

\begin{figure}[t!]
\centering
        \includegraphics[scale=0.24]{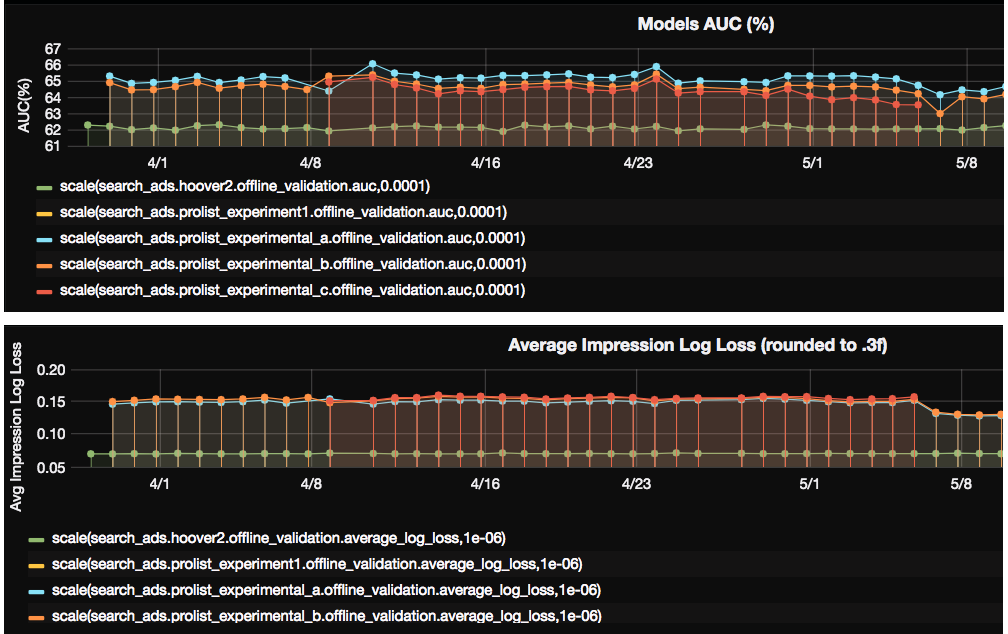}
\caption{Offline evaluation metrics: {\tt AUC} and Average Impression Log Loss on the hold out data (one day of promoted listings search click logs)}
\label{fig:metrics1}
\end{figure}

\medskip

\begin{figure}[t!]
\centering
        \includegraphics[scale=0.24]{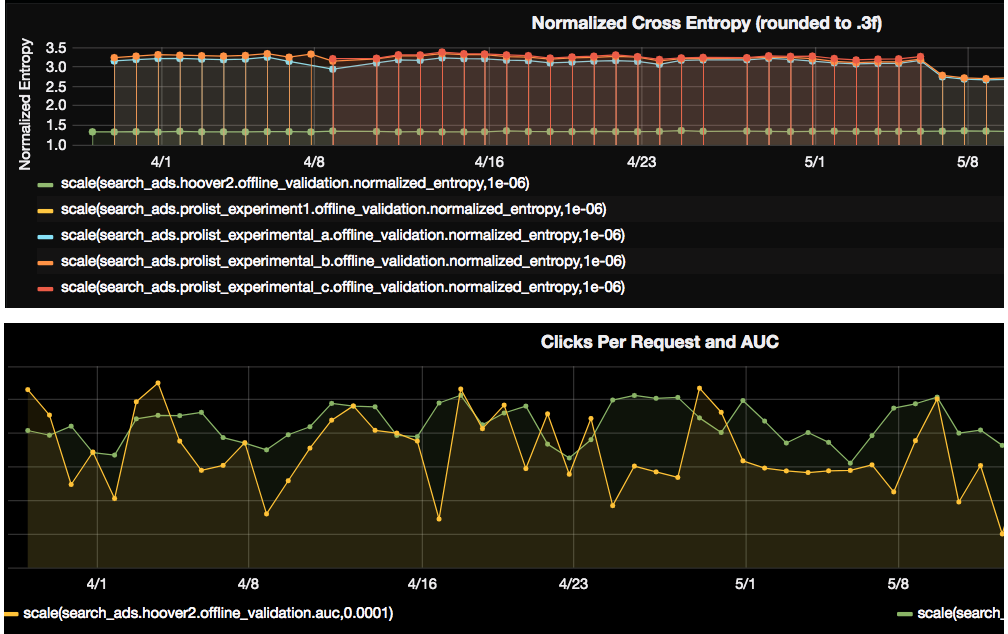}
       
\caption{Offline evaluation metrics: Normalized cross entropy on the hold out data (one day of promoted listing search click logs) and overlayed {\tt AUC} with clicks per request per day}
\label{fig:metrics2}
\end{figure}

{\bfseries AUC}: {\tt AUC} is a good metric for measuring ranking quality without accounting for calibration~\cite{He2014}. Empirically, we have found {\tt AUC} to be a very reliable measure of relative predictive power of our models. Improvements in {\tt AUC} ($> 1\%$) have consistently resulted in significant {\tt CTR} wins in multiple rounds of online A/B tests. Figure~\ref{fig:metrics1} shows offline {\tt AUC} of multiple models. Figure~\ref{fig:metrics2} (bottom figure) shows {\tt AUC} and clicks per request for a production model.

\medskip

{\bfseries Log Loss}: Log loss is the negative log-likelihood of the Bernoulli model and is often used as an indicator of model performance in online advertisement. Minimizing log loss indicates that $P(c\, | \, q,l)$ should converge to the expected click rate and result in a better model~\ref{fig:metrics1}. A model with lower average impression log loss is then preferred. Figure~\ref{fig:metrics1} (bottom figure) shows the average impression log loss for multiple variants compared to a baseline model.

\medskip

{\bfseries Normalized Cross Entropy}: This metric is proposed in He et al.~\cite{He2014}. In essence, normalized cross entropy is the average impression log loss normalized by the entropy of average empirical {\tt CTR} of the training dataset. In each training of the model, we compute the average empirical {\tt CTR} for the training set used by each variant. We then divide the average impression log loss by this empirical training CTR to obtain the normalized cross entropy. The key benefit normalized cross entropy is its robustness to empirical training CTR. Figure~\ref{fig:metrics2} (top figure) shows this metric for multiple variants compared to a baseline model.

\subsection{Empirical Results}\label{sec:exp_res}
Here, we present and discuss the results of our offline evaluations on metrics discussed in Section~\ref{sec:exp_metrics} for a typical date. Recall that for our ensemble learning described in Section \ref{sec:method_ensemble}, we want final higher level models to learn how to balance different individual models in different data regions. Based on some of our preliminary experiments, we found that simply dividing datasets based on the number of impressions are already a good starting point. In particular, in this experiment, the training and test sets are both divided into two partitions: listings with more than or equal to $k$ impressions during the training period (denoted as \textit{warm}) and listings with less than $k$ impressions during the training period (denoted as \textit{cold}). We denote the original training and testing dataset as \textit{mixed} datasets. We set $k=30$ empirically as the impression breakdown threshold for cold and warm datasets. Here, we only report offline experimental results, but the reality that online A/B tests share  similar results as {\tt AUC} is a consistent indicator of the offline/online experiments correlation.

\newcommand{\ra}[1]{\renewcommand{\arraystretch}{#1}}
\begin{table*}\centering
\ra{1.2}
\begin{tabular}{@{}rrrrcrrrcrrr@{}}\toprule
& \multicolumn{3}{c}{Historical} & \phantom{abc}& \multicolumn{3}{c}{Content-based} &
\phantom{abc} & \multicolumn{3}{c}{Ensemble}\\
\cmidrule{2-4} \cmidrule{6-8} \cmidrule{10-12}
& mixed & cold & warm && mixed & cold & warm && mixed & cold & warm \\ \midrule
AUC ($\%$)\\
& $+1.56$ & $+1.89$  & $+1.55$ && $-1.57$ & $+6.39$ & $-1.74$ && $\mathbf{+1.95}$ & $\mathbf{+8.34}$ & $\mathbf{+1.83}$\\
Log Loss ($\times10^3$)\\
& $-0.016$ & $-0.048$ & $-0.018$ && $+0.311$ & $-0.194$ & $+0.335$ && $\mathbf{-0.092}$ & $\mathbf{-0.332}$ & $\mathbf{-0.087}$\\
Normalized Entropy ($\times10^3$)\\
& $-0.29$ & $-1.23$ & $-0.31$ && $+5.67$ & $-5.00$ & $+6.01$ && $\mathbf{-1.68}$ & $\mathbf{-8.55}$ & $\mathbf{-1.56}$\\
\bottomrule
\end{tabular}
\vspace{5mm}
\caption{Changes in {\tt AUC} ($\%$), average impression log loss ($\times 10^3$) and normalized cross entropy ($\times10^3$) on the dataset is compared to a numerical listing id only model, which was deployed as the production model, across different variants. The historical model utilized historical smoothed  {\tt CTR} for the logistic regression while the content-based model uses the multimodal listing features discussed in Section~\ref{sec:method_features}. The ensemble model is trained with content-based and historical models scores along with $\lfloor\log (impression_{count})\rfloor$ as an input to the ensemble. The impressions break threshold for this experiment is set at $k=30$ as discussed in Section~\ref{sec:exp_res}.}
\label{tab:results}
\end{table*}

Table~\ref{tab:results} summarizes the results of our experiments in terms of {\tt AUC}, average impression log loss and normalized cross entropy. All models are compared to a baseline model: a logistic regression with {\tt FTRL} trained on \textit{mixed} training dataset using a numerical listing ID as the single feature representation, which was deployed as the production model. We report the changes in metrics compared to this baseline. The \textbf{historical} model denotes the model trained on the \textit{warm} training dataset with average smoothed {\tt CTR} of Section~\ref{sec:method}as the single listing representation. The \textit{cold}, \textit{warm}, and \textit{mixed} metrics show the relative performance of this model in comparison with the baseline model. We can observe that the historical model outperforms the baseline in terms of evaluation metrics on \textit{warm}, \textit{cold} and \textit{mixed} testing sets.

The \textbf{content-based} model is a model that utilizes a multimodal (text and image) feature representation of listings in the training dataset. The content-based model performs significantly better on the \textit{cold} testing set as illustrated in Table~\ref{tab:results}. As expected this model performs worse in terms of offline evaluation metrics on \textit{warm} and \textit{mixed} training sets.

The \textbf{ensemble} model takes the previous models' predictions (historical model trained on \textit{warm} training set and content-based model trained on \textit{cold} to maximize its performance) along with smoothed impressions ($\lfloor\log (impression_{count})\rfloor$) as the input feature. This model performs significantly better in terms of {\tt AUC} ($> 1.9\%$ lift on the mixed testing set, $> +8\%$ lift on cold testing set, and $>+1.8\%$ lift on the warm testing set), log loss, and normalized entropy on the mixed testing set. The model also performs significantly better compared to the baseline and previous models on \textit{cold} and \textit{warm} testing sets.


\section{Conclusion}\label{sec:conclusion}
In this paper, we presented an overview of how promoted listings' {\tt CTR} prediction system works at Etsy. In addition to the holistic view, we proposed an ensemble learning approach to leverage different signals of listings. In order to learn meaningful representations from features, we utilized feature hashing and deep learning techniques to obtain representations from texts and images and employ a multimodal learning process to combine these different signals. To demonstrate the effectiveness of the system, we compared the proposed system to non-trivial baselines on real world data from Etsy in the offline setting, which correlates online experimental results. During this process, we also established a strong correlation between offline evaluation metrics and online performance ones, which serves as a guidance for future experimental plans. This paper serves as the first study to systematically discuss how a promoted listings system can be built with practical considerations. 

\section{Acknowledgment}\label{sec:ack}
We gratefully acknowledge the contributions of the following: Corey Lynch, Alaa Awad, Stephen Murtagh, Rob Forgione, Jason Wain, Andrew Stanton, Aakash Sabharwal, and Manju Rajashekhar.

\bibliographystyle{ACM-Reference-Format}
\bibliography{adkdd} 

\end{document}